\documentclass[a4paper,11pt]{article}
\usepackage[switch]{lineno} 
\pdfoutput=1 

\usepackage{jinstpub} 

\usepackage{listings} 
\usepackage{caption}
\usepackage{microtype}
\usepackage{siunitx}
\usepackage{cleveref}
\usepackage{subcaption}
\usepackage{amsmath}
\usepackage{amssymb}
\usepackage{amsthm}
\usepackage{gensymb}
\usepackage{mathtools}
\usepackage{algorithm}
\usepackage{algpseudocode}
\usepackage{newtxtext,newtxmath,amsmath}

\newcommand{\argmin}[1]{\underset{#1}{\operatorname{arg}\,\operatorname{min}}\;}

\title{\boldmath An evolutionary strategy for $\Delta$E - E identification}

\author[a,b,1]{K. Schmidt,\note{Corresponding author.}}
\author[c]{O. Wyszy\'nski}

\affiliation[a]{Institute of Physics, University of Silesia,\\Katowice, Poland}
\affiliation[b]{Cyclotron Institute, Texas A$\&$M University\\College Station, TX, USA}
\affiliation[c]{Faculty of Physics, Astronomy and Applied Computer Science,\\Jagiellonian University, \L{}ojasiewicza 11, 30-348 Krak\'ow, Poland}

\emailAdd{Katarzyna.Schmidt@us.edu.pl}

\abstract{
In this article we present an automatic method for charge and mass identification of
charged nuclear fragments produced in heavy ion collisions at intermediate
energies.  The algorithm combines a generative model of $\Delta$E - E relation
and a Covariance Matrix Adaptation Evolutionary Strategy (CMA-ES).  The CMA-ES
is a stochastic and derivative-free method employed to search parameter space of
the model by means of a fitness function.  The article describes details of the
method along with results of an application on simulated labeled data.
}

\keywords{particle identification methods, heavy-ion, detector, Pattern recognition, 
calibration, fitting, CMA-ES, covariance matrix adaptation, evolutionary strategy}

\arxivnumber{1234.56789} 

\begin{document}
\maketitle
\flushbottom

\section{Introduction}
\label{sec:intro}
\begin{figure}[H]
  \centering
  \includegraphics[width=0.5\textwidth]{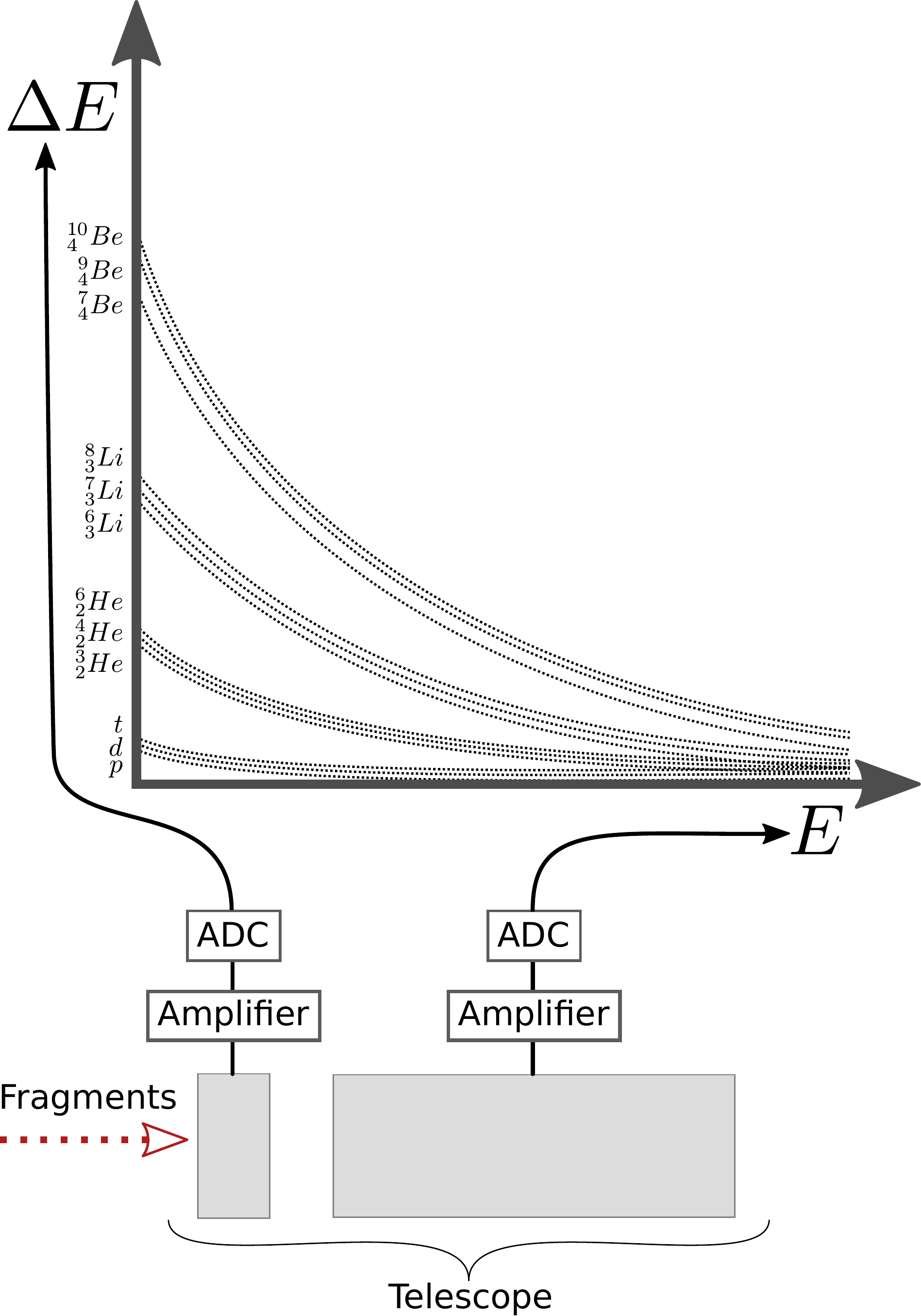}%
  \caption{A schematic overview of $\Delta$E-E identification procedure. 
  Different fragments produced in a nuclear reaction and stopped in a 
  telescope, populate identification lines characteristic of their charge and mass.}
  \label{fig:overwiev}
\end{figure}

In heavy ion collisions at intermediate energies many fragments of different
mass (A) and charge (Z) are produced due to multi-fragmentation processes.
The number of those fragments is not constant and depends on many conditions: mass
of target and projectile, impact parameter and the energy of the projectile. The
fragments are often detected in telescopes, stacks of different thickness
detectors, assembled in an array to cover the full solid angle.  Such
detection systems have been developed in many research institutes
\cite{stracener1990dwarf, pouthas1995indra, kwiatkowski1995indiana,
sarantites1996microball, pagano2004fragmentation, wuenschel2009nimrod,
bougault2014fazia, lukasik2013kratta}.  The number of telescopes in such arrays
vary from few hundreds \cite{wuenschel2009nimrod} to over a thousand
\cite{pagano2004fragmentation}.  The measurements of fragment energy losses in
each layer of the telescope are the key to the identification process.  The
correlation between the energy loss in one or several layers versus the residual
energy released in the detector in which the particle has stopped, reveals
specific curved lines, each representing a set of signals from fragments of the same A
and Z (see \cref{fig:overwiev}). 
The general performance of the detectors depends obviously on their quality and
the associated electronics but also on the homogeneity of their response and
their stability over long periods of time (e.g. temperature). Thus, each telescope can 
give a different $\Delta$E-E matrix, which can change
during data-taking. 
In the end, the identification procedure has
to be performed several times for each telescope before physics analysis
can start.\\ Nowadays there are several methods used to determine mass and
charge of detected fragments.  In general, they can be grouped in several
classes, however some parts of those methods may overlap.
\begin{enumerate}
  \item Graphical - is the most trivial set of methods where an user draws 
  interactively and by hand curves on the $\Delta$E-E matrix.
  Particles are then identified by comparing distances to the closest lines.
  \item Fitting methods:
    \begin{enumerate}
      \item Algorithms are used to find parameters of a function describing 
        $\Delta$E-E correlation for a subset of ridges\cite{butler1970x, tassan2002new, le2002mass}.  
        Particle identification is obtained by
        inversion of the function for given $\Delta$E and E, in order to extract
        Z and possibly A.
      \item Methods that calculate energy loss tables with the use of a priori knowledge about the 
            incident ion (mass, charge and energy) and about the absorber medium 
            (volumetric density and atomic number) ~\cite{mastinu1996procedure,
            dudouet2013comparison,lukasik2013kratta,cap2013detection}. 
            Each particle is then identified from its relative
            distance between pairs of the closest ridge lines.
    \end{enumerate}
  \item Peak finding methods:
     \begin{enumerate}
       \item Methods built on top of one of the above methods, the lines need to be
         drawn by hand or fitted to the $\Delta$E-E matrix, which is followed by 
         the linearization procedure~\cite{wuenschel2009nimrod}.  After linearization, data is projected
         onto a one-dimensional plot producing quasi-Gaussian peaks.
         The isotopic peaks within an element are
         fitted with Gaussian functions that are then correlated to masses.
       \item A part of the matrix is projected onto a relevant helper line, D($\theta$),
         followed by a peak localization. 
         The operation of projection/localization is then repeated in order to
         cover the full matrix, varying a helper line $\theta$ from $0$ to
         $90\degree$. \cite{gruyer2016new}
         
     \end{enumerate}
\end{enumerate}
The quality of the identification procedure in the first, graphical method is fully
user dependent and extremely time consuming. It does not allow any extrapolation
into the low statistics region, which is possible in other described
methods. Fitting procedures are very sensitive to statistics and
they usually require additional constraints provided by a user. 
For example in method described in 2(b) it is extremely important to know with good precision each $\Delta$E 
detector the active thickness and dead layer. 
The last mentioned method is the least user-intervention dependent ("only two
mouse-clicks from the user to calculate all initialization parameters"), however
the choice of the first click seems to be crucial for a final result. 
The histogram binning also has to be chosen carefully. 
The strength of the method is that it can be used to
many types of 2 dimensional matrices, instead of only $\Delta$E-E correlations.
Nevertheless, the common feature of all above methods is that they require a
dedicated graphical interface. The method presented in this article 
is an adaptation of evolutionary algorithm for particle trajectory 
reconstruction~\cite{wyszynski2016evolutionary}.
It belongs to a class of fitting methods, however it provides great improvements.
It does not require any user interaction, any initial parameters provided and also does not require a
graphical interface. It requires only a file with the x and y positions
of points being signals from the E and $\Delta$E detectors and a model, which
describes the relation between those points. As the result of the algorithm, the file
gains two more columns, which are Z and A assigned to each E and $\Delta$E pair.
  In the following sections, we describe the model used to interpret data
as well as the data association and the classification algorithms.
At the end, a test procedure is described and the results are presented together
with short discussion on strong points as well as drawbacks.

\section{The model \label{sec:model}}
Correlations between measured energy losses in two successive detectors ($\Delta$E - E)
create a specific pattern, presented on \cref{fig:inputdata}, which can be
easily modeled.
In order to build a model, we have used the function proposed by L.Tassan-Got~\cite{tassan2002new}. 
For detectors delivering a linear response (e.g. silicon detectors), it reads:
\begin{equation}
\label{eq:TG1}
\Delta E = t(E,g,\mu,\lambda,A,Z) = \Bigg[(gE)^{\mu + 1} + \Big(\lambda Z^{\frac{2}{\mu + 1}} A^{\frac{\mu}{\mu + 1}}\Big)^{\mu + 1}\Bigg]^{\frac{1}{\mu + 1}} - gE
\end{equation}
where the parameters we are looking for are G, $\lambda$ and $\mu$.
However it has already been noticed in the past~\cite{butler1970x,shimoda1979simple} 
that for a wider Z range, the above formula must be extended by additional
parameters $\alpha$, $\beta$, $\nu$ and $\xi$: 
\begin{equation}
\label{eq:TG2}
\Delta E = t(E,g,\mu,\nu,\lambda,\alpha,\beta,\xi,A,Z) = 
\Bigg[ (gE)^{\mu + \nu + 1} + \Big(\lambda Z^{\alpha} A^{\beta}\Big)^{\mu + \nu + 1} + \xi Z^{2}A^{\mu}(gE)^{\nu} \Bigg]^{\frac{1}{\mu + \nu + 1}} -gE
\end{equation}
For a detector delivering a non linear response versus deposited energy this
function needs to be corrected. The energy \textit{E} in equation~\ref{eq:TG2}
must now be expressed as a function of the light \textit{h} emitted by the
scintillator. The light response of CsI(Tl) crystals is deduced from the Birks
formula~\cite{birks1965theory} and allows expression of the energy released in the
CsI as a function of the emitted light :
\begin{equation}
\label{eq:Eh}
E = \sqrt{h^{2} + 2\rho h \Bigg[ 1 + \ln{\big(1 + \frac{h}{\rho}\big)}\Bigg]} 
\end{equation}
where $\rho = \eta Z^{2}A$ is a new parameter.\\
In our method we have used the function for linear detectors (\ref{eq:TG1}) to
build our model in the following manner:
\begin{equation}
\label{eq:model}
  \mathrm{model} \quad
  \begin{cases}
  t(E,g,\mu,\lambda,A_1,Z_1)\\
  \dots\\
  t(E,g,\mu,\lambda,A_n,Z_n)\\
  \end{cases}
  \quad \mathrm{where} \quad \{A_i,Z_i\} \in I 
\end{equation}
The set $I$ contains mass number (A) and atomic number (Z) pairs for isotopes of
interest, in the presented model from $^{1}_{1}$H up to $^{25}_{12}$Mg.
The model is easily adjustable with regards to the type of fragments by manipulating
the parameters $A$ and $Z$. When model is ready, the next step is to interpret
the data.
In the following section, we describe the data to model association algorithm 
as well as the method used to search the model parameter space.

\section{Data classification \label{sec:method}}
The algorithm is based on a generative model of $\Delta$E - E correlation. 
The model is iteratively compared with input data by means of a user defined
fitness function. The function compares data and the model (\cref{eq:model}) 
with given parameters in order to evaluate the correctness of model parameters.
In the evolutionary strategy the definition of a fitness function is a key point. 
We define it as follows:
\begin{equation}
\label{eq:fitness}
f(D_{m,2}\,,g,\mu,\lambda) = \sum_{j=1}^{m} \argmin{i \in
I} (D_{j,1} - t(D_{j,2}\,, g, \mu, \lambda, A_i, Z_i))^2
\end{equation}
where
\begin{equation}
\label{eq:datamatrix}
D_{m,2} = 
  \begin{bmatrix}
      \Delta E_1  & E_1 \\
      \dots  &  \dots\\
      \Delta E_m  & E_m \\
  \end{bmatrix}
\end{equation}
denotes a matrix of $m$ input data points ($\Delta E$ and $E$ of fragments). 
In other words, the fitness function $f(\dots)$ 
quantifies minimal residual between every data point $p_j = [D_{j,1}, D_{j,2}]$
and its closest function $t(\dots, A_i,Z_i)$ of the model.

With the model and fitness functions defined, the parameters
$g$, $\mu$, $\lambda$ have to be estimated.
In order to search the model parameter space, we have employed a method called Covariance Matrix
Adaptation Evolutionary Strategy (CMA-ES) \cite{hansen2006eda, hansen1996IEEE,
hansen2001ecj, igel2007ecj}.

The CMA-ES is an iterative, stochastic and derivative-free method designed for
difficult non-linear, non-convex, black-box optimization problems in the continuous domain. 
This method is considered as state-of-the-art in evolutionary computation and 
it is typically applied to unconstrained optimization problems with 
search space dimensions between three and a hundred. 
The CMA-ES is an alternative for quasi-Newton methods such as very popular
Broyden-Fletcher-Goldfarb-Shanno (BFGS) algorithm, in cases where
derivatives are not available.

We have implemented the method using the R language\cite{team2000r}
and C++ CMA-ES library\cite{benazera2014libcmaes} integrated
by means of Rcpp package\cite{eddelbuettel2011rcpp}.
The decision to choose C++ implementation of CMA-ES has been made 
based on execution speed (it supports multi-threading)
as well as due to advanced development stage of the project. 
The library is released under the GPLv3 license and offers rich set of options 
such as possibility to define basic CMA-ES parameters (e.g. $\mu$, $\lambda$), 
model parameter range bounds, gradient function, various top criteria,
variation of CMA-ES method and more.

In the CMA Evolution Strategy, a population of new search point is generated 
by sampling a multivariate normal distribution $\mathbb{R}^{n}$.
Those points are evaluated using a fitness function in order to select
the best fit candidates. Afterwards, a new weighted mean is calculated 
based on selected candidates. The covariance matrix is adjusted accordingly
in order to update the space search region for next iteration.
The method described in this article uses a variation of CMA-ES called
Active-CMA-ES\cite{aCMAES2006}, or in short, aCMA-ES . 
It differs from the regular one by selecting 
not only the best candidates but also the worst ones.
The latter are used to adapt a covariance matrix faster towards minimum value
of objective function. The aCMA-ES has proven to outperform classical variations
of the algorithm\cite{hansen2010benchmarking}.

After the aCMA-ES algorithm finds the minimum of objective
function, the model parameters are used to classify the data.
The classification is performed based on distance 
\begin{equation}
\label{eq:dist}
d = |D_{j,1} - t(D_{j,2}\,,g,\mu,\lambda,A_i,Z_i)|
\end{equation}
between a data point $p_j$ and a function $t(\dots, A_i,Z_i)$ of the model.
For example, if a value of the function $t(E_j,\dots, A_{25}, Z_{12})$
is the closest to a data point $p_j$, 
then the point $p_j$ is classified as $^{25}_{12}$Mg.

In the following section, we present the performance of the working algorithm
on simulated data with superimposed Gaussian noise.

\section{Results \label{sec:results}}
\begin{figure}[H]
  \centering
  \includegraphics[keepaspectratio, page=1, width=0.5\textwidth]{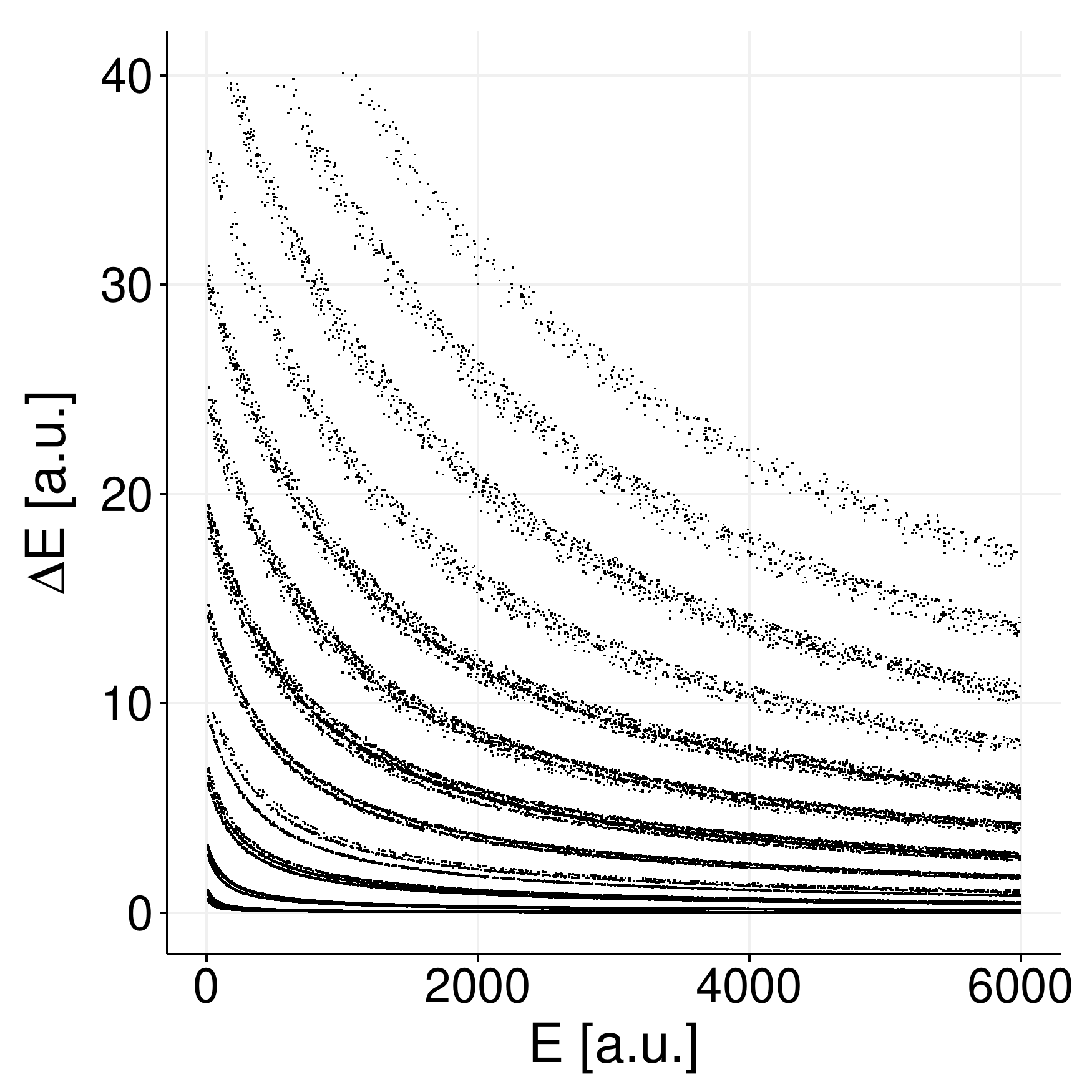}%
  \includegraphics[keepaspectratio, page=3, width=0.5\textwidth]{EffBarPlot_reference.pdf}
  \caption{Left: $\Delta$E - E points simulated using eq.~\ref{eq:TG1} with Gaussian noise. Numbers of particular isotopes have been 
          based on real telescope from NIMROD array. Right: Identification discrepancies of particular fragments after noise application.
          Negative values indicate number of fragments that have been
          underestimated, whereas positive indicate that they have been overestimated.}
  \label{fig:inputdata}
\end{figure}

\begin{wrapfigure}{l}{0.36\textwidth} 
      \tiny
      \begin{minipage}[b]{\linewidth}
        \centering
        \includegraphics[keepaspectratio, page=4, width=\textwidth]{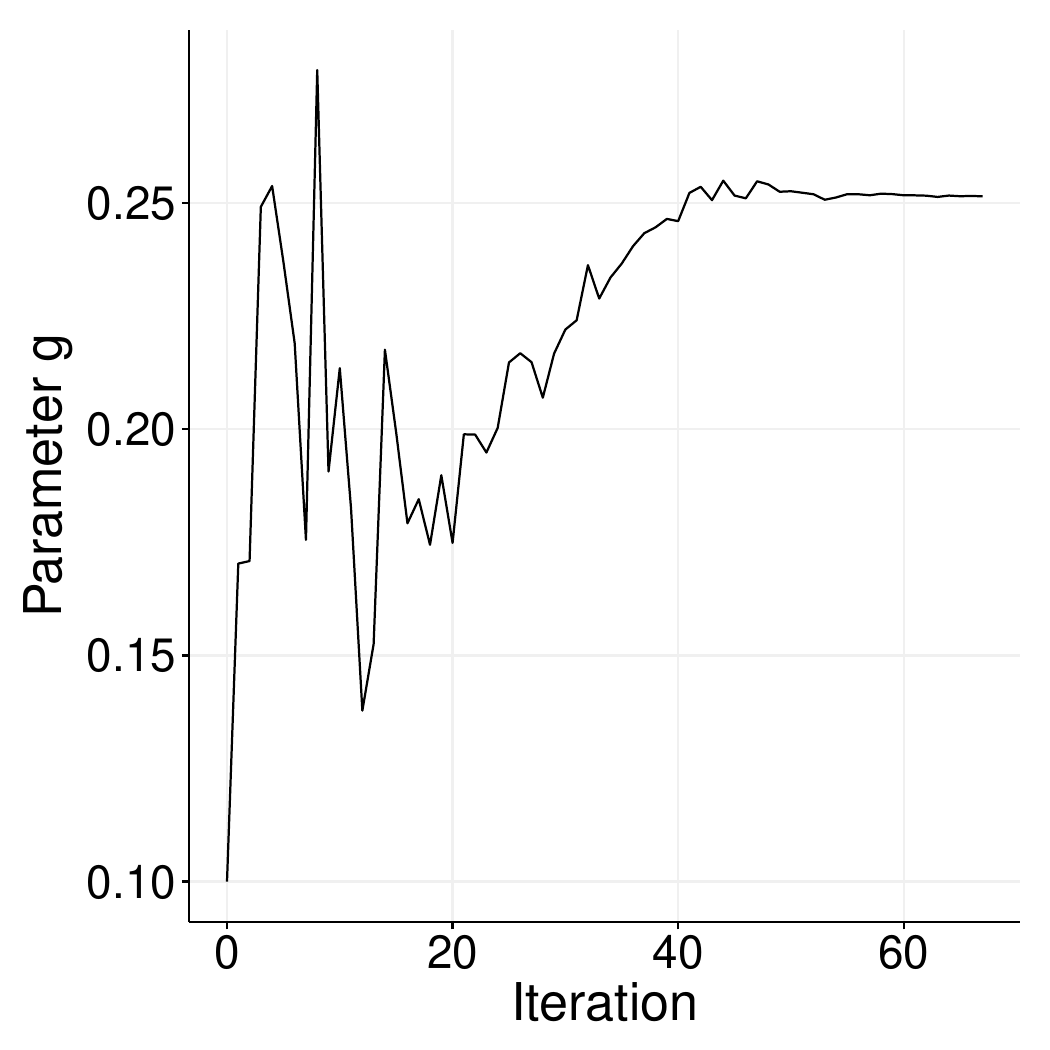}
        \subcaption{}
        \label{fig:progress4}
      \end{minipage}
      \begin{minipage}[b]{\linewidth}
        \centering
        \includegraphics[keepaspectratio, page=5, width=\textwidth]{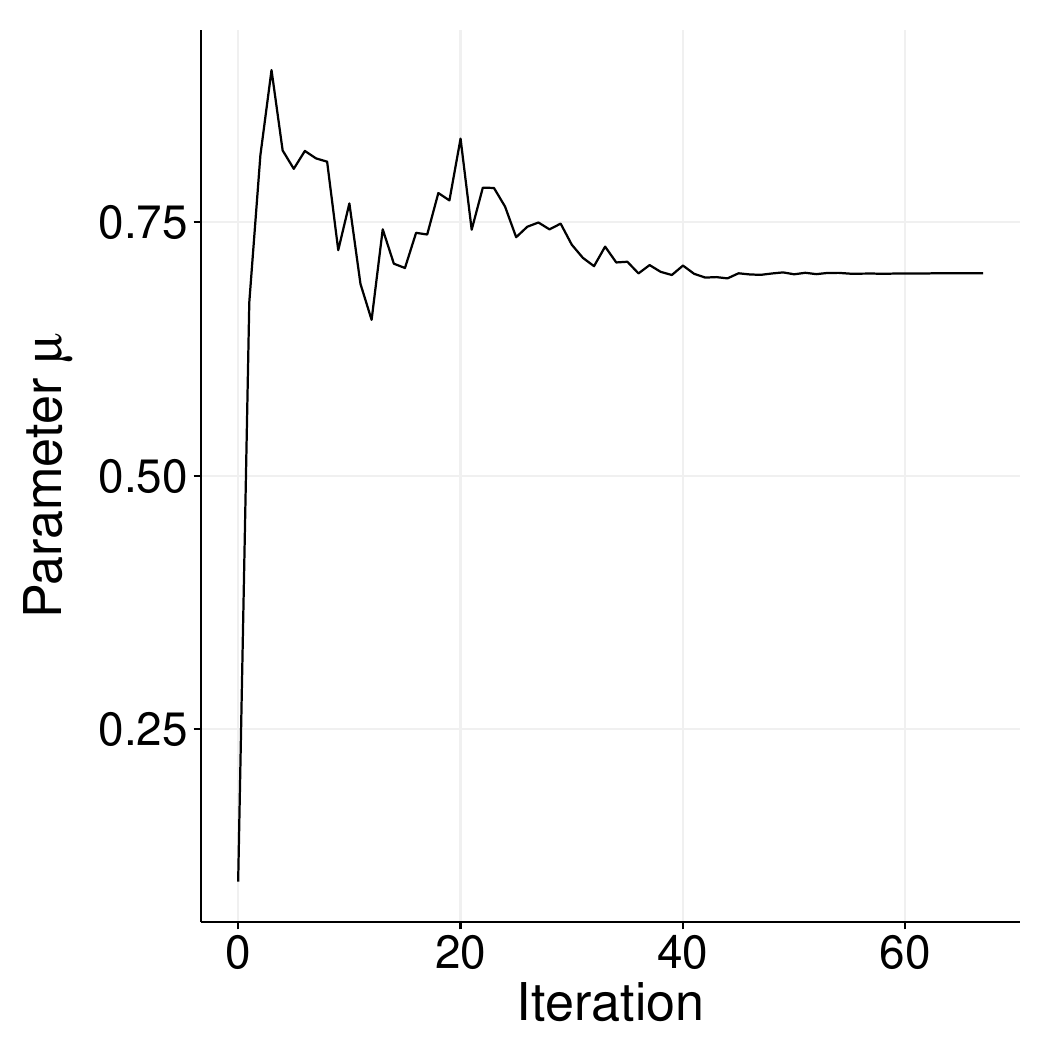}
        \subcaption{}
        \label{fig:progress5}
      \end{minipage}
      \begin{minipage}[b]{\linewidth}
        \centering
        \includegraphics[keepaspectratio, page=6, width=\textwidth]{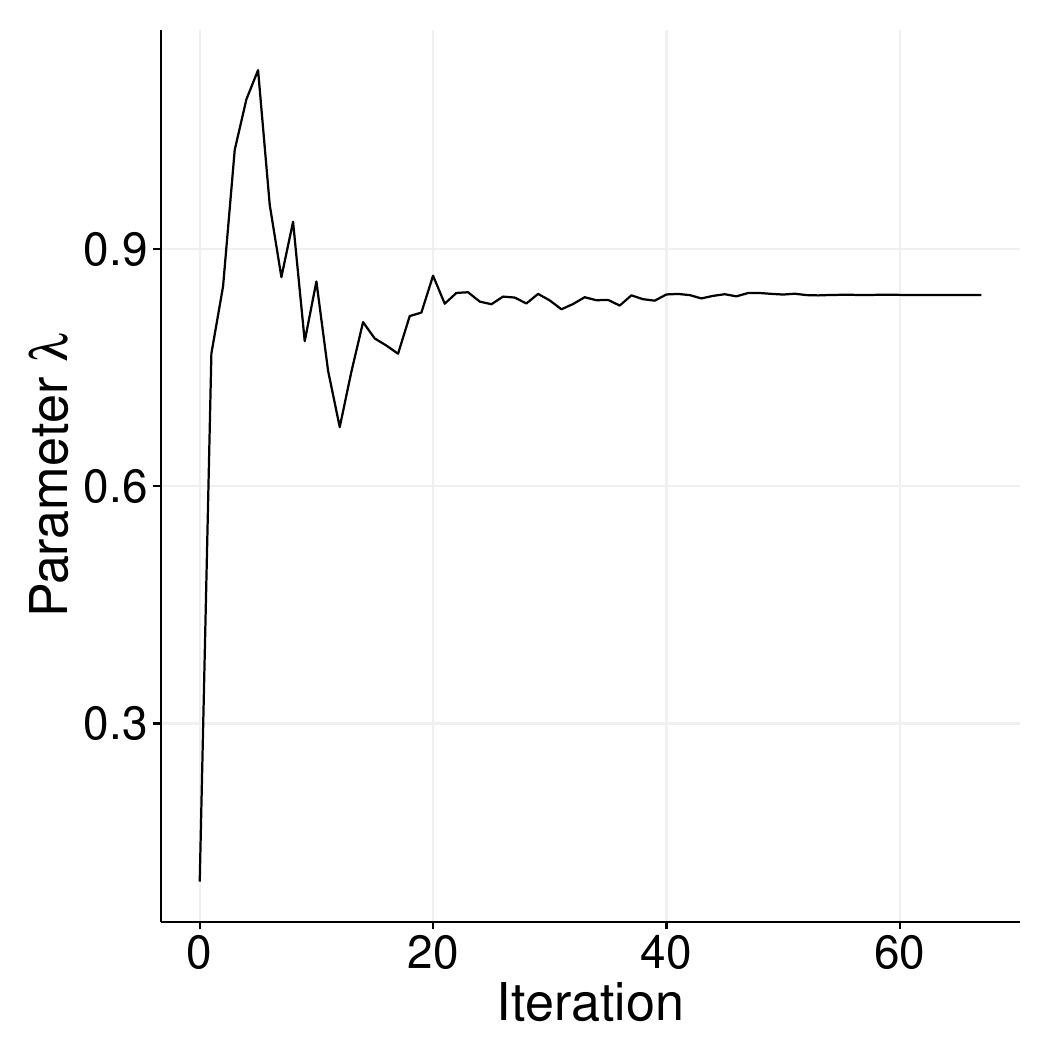}
        \subcaption{}
        \label{fig:progress6}
      \end{minipage}
      \caption{Model function parameters evolution: (a) parameter g, (b) parameter $\mu$, (c) parameter $\lambda$}
      \label{fig:cmaesprogress}
      \vspace{-1cm}
\end{wrapfigure}

\begin{figure}[ht]
      \tiny
      \begin{minipage}[b]{.5\linewidth}
        \centering
        \includegraphics[keepaspectratio, width=\textwidth]{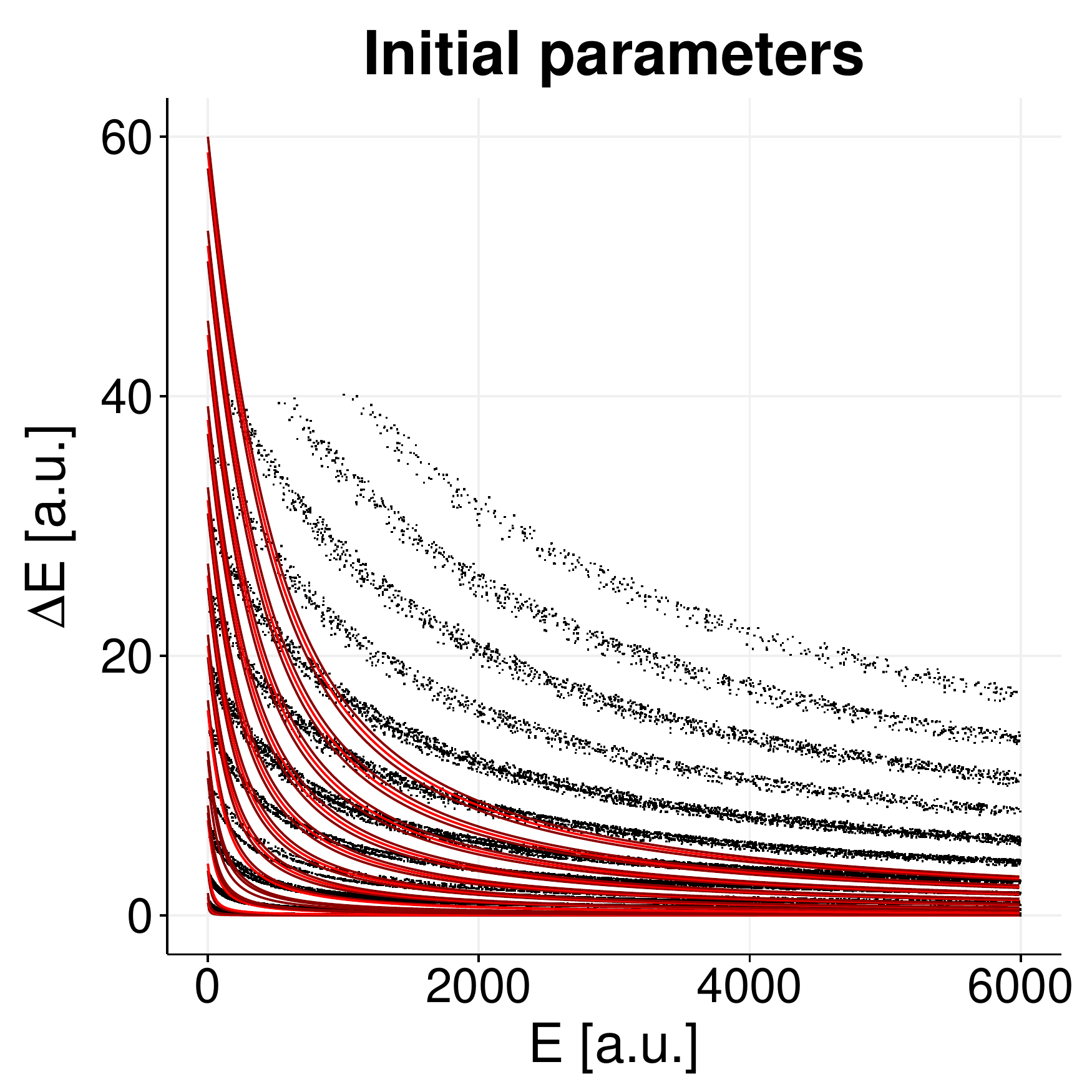}
        \subcaption{}
        \label{fig:modelevo1}
      \end{minipage}%
      \begin{minipage}[b]{.5\linewidth}
        \centering
        \includegraphics[keepaspectratio, width=\textwidth]{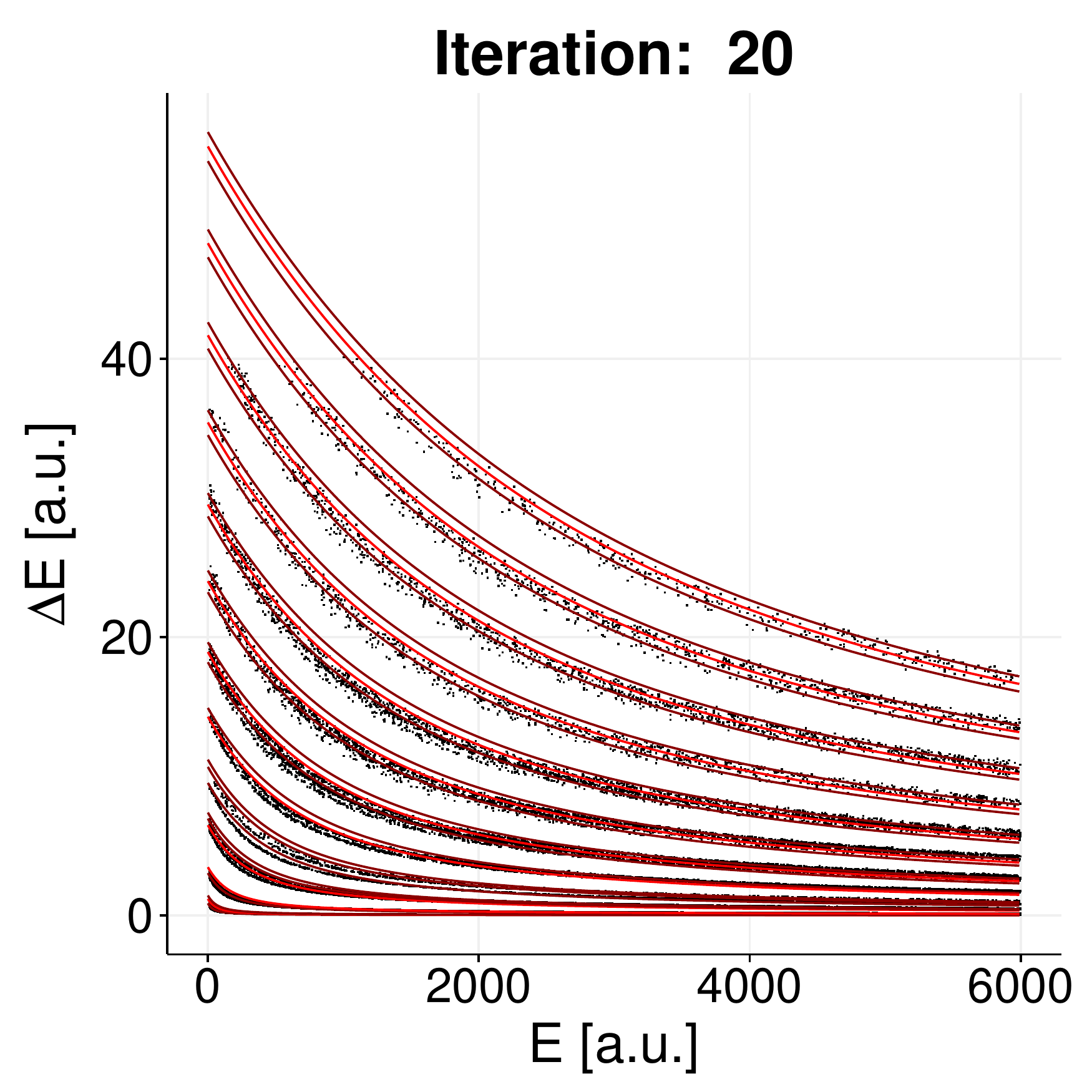}
        \subcaption{}
        \label{fig:modelevo2}
      \end{minipage}
      \begin{minipage}[b]{.5\linewidth}
        \centering
        \includegraphics[keepaspectratio, width=\textwidth]{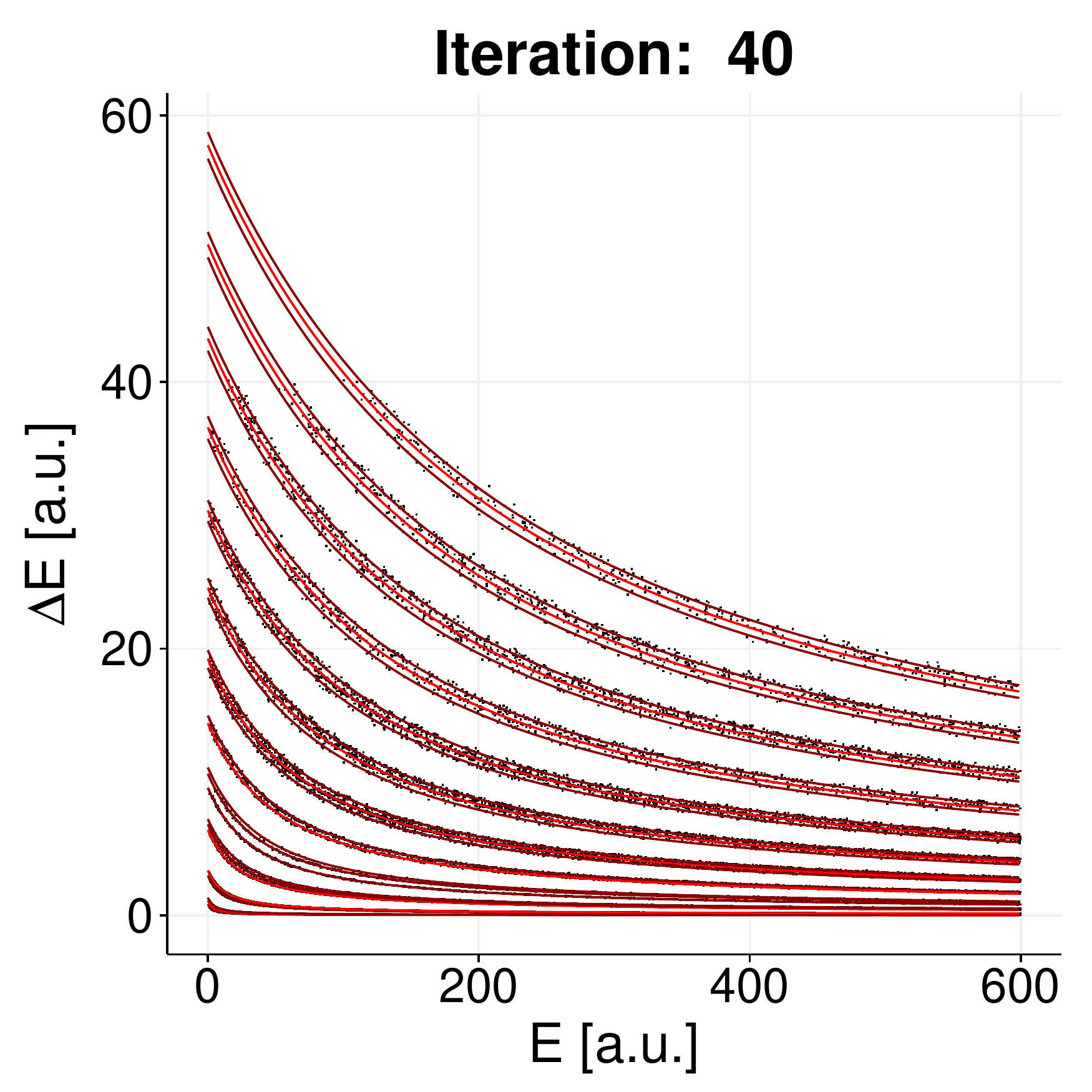}
        \subcaption{}
        \label{fig:modelevo4}
      \end{minipage}%
      \begin{minipage}[b]{.5\linewidth}
        \centering
        \includegraphics[keepaspectratio, width=\textwidth]{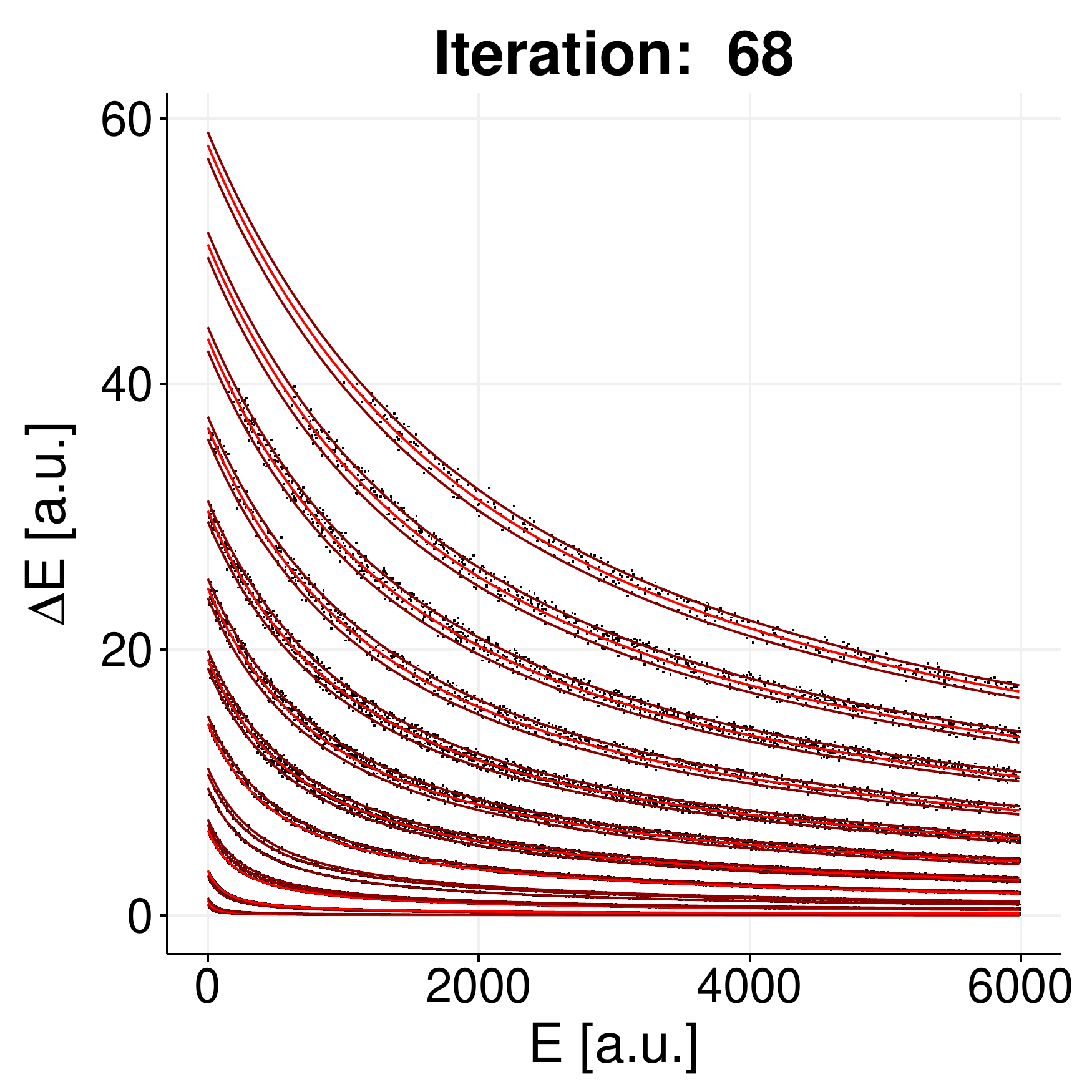}
        \subcaption{}
        \label{fig:modelevo6}
      \end{minipage}
      \caption{The model function (red lines) calculated with the initial parameters (a), after the 20th (b), 40th (c) and 68th (d) iteration, plotted on the 
               top of simulated data (black points).}
      \label{fig:modelevolution}
      \vspace{-1cm}
    \end{figure}

\begin{figure}[ht]
      \tiny
      \centering
      \includegraphics[keepaspectratio, page=3, width=0.7\textwidth]{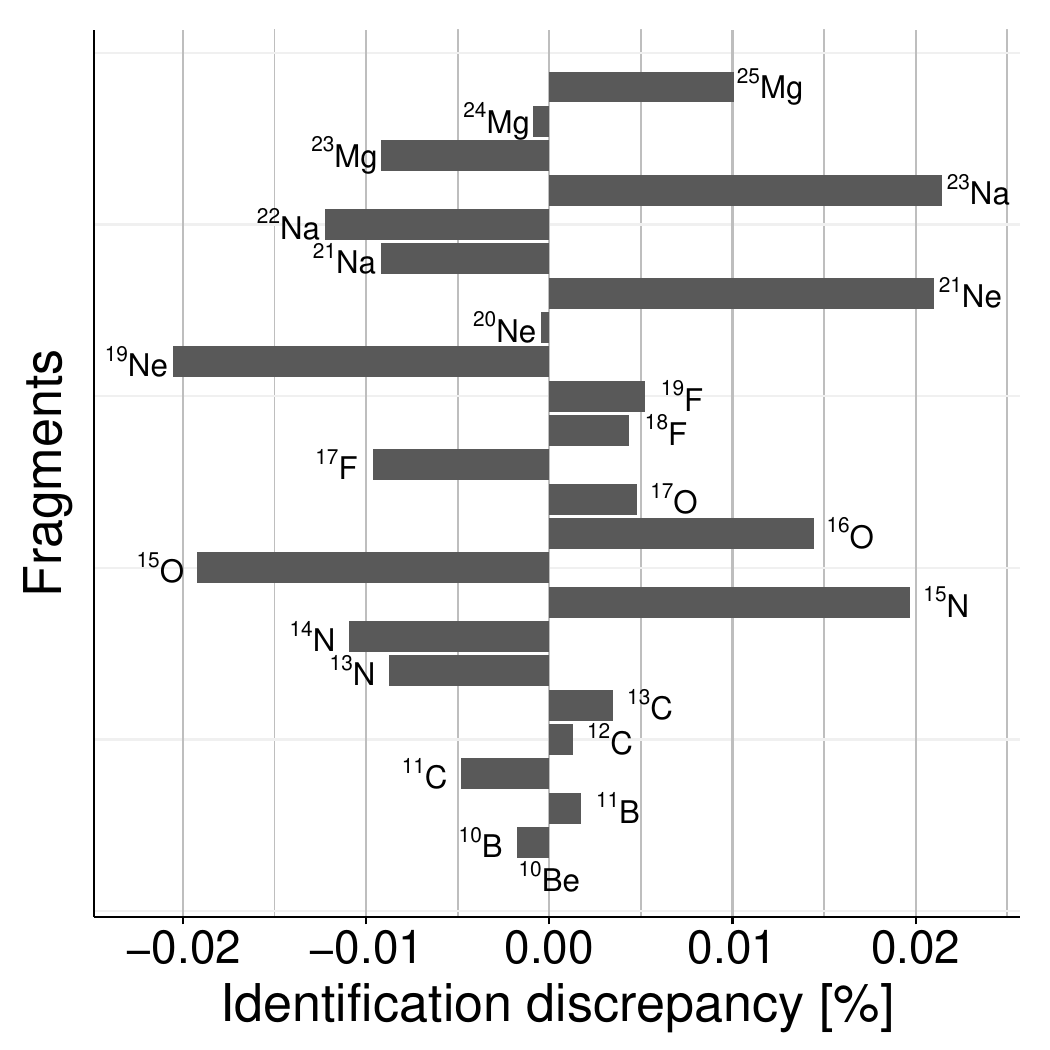}
      \caption{Identification discrepancies of particular fragments after final identification process.. 
               Negative values inform that numbers of fragments have been underestimated, positive - that they have been overestimated.}
      \label{fig:eff_result}
    \end{figure}

In order to test the algorithm, labeled data has been simulated with
superimposed Gaussian noise as presented on the left panel of
\cref{fig:inputdata}. The numbers of particular isotopes (230k in total) have
been based on real telescope from the NIMROD array~\cite{wuenschel2009nimrod}.  The
data was generated using set of functions described by \cref{eq:model}, which
define our generative model.
The parameters of the model g, $\mu$ and $\lambda$ were 0.25, 0.7 and 84,
respectively.  As a first step, the re-classification of data to lines generated
with these parameters was performed.  Due to a noise application, some of the
$\Delta$E - E points have changed their position with respect to the original
one so strongly, that their mass classification failed for 501 (0.22$\%$)
fragments with Z > 4. This means, that if our evolutionary algorithm
reconstructed the model function parameters with 100$\%$ accuracy, 501 fragments
would be in any case not correctly classified in mass number.  The atomic number
has been assigned correctly to all studied fragments.  In the right panel of
\cref{fig:inputdata} we present the identification discrepancy for each fragment due
to a noise application. We calculate it as a difference between the originally
simulated number of fragments of given A and Z and number of these fragments
after noise application and re-classification, normalized to the total number of
simulated fragments.  Negative values indicate number of fragments that have been
underestimated, whereas positive indicate that they have been overestimated.  The fragments
with A and Z reassigned to simulated data after noise superimposition, will be,
from now on, called the reference data.
The evolutionary algorithm has been run on an average class laptop. It needed
around 10 minutes and 68 iterations to complete the calculations.  The initial
values of parameters g, $\mu$ and $\lambda$ have all been set to 0.1. Since the
aCMA-ES works better when the parameters are the same order of magnitude,
parameter $\lambda$ has been scaled by 100. As a result of the algorithm, they
have been estimated to 0.252, 0.699 and 0.842, respectively. In
\cref{fig:cmaesprogress} we show the evolution of those parameters as a function
of iteration number.  The model function calculated with the initial parameters
(a), after the 20th (b), 40th (c) and 68th (d) iteration, plotted on the top of
simulated data is shown in \cref{fig:modelevolution}. It can be seen, that the
model function with initial parameters certainly does not describe the simulated
data, however after the 20th iteration it decently fits the data. The last two
thirds of the iterations are used to fine-tune the parameters.
Parameters obtained as the result of the evolutionary algorithm have been used
to produce a final mass and charge classification. Fragments with A and Z assigned to
simulated data after algorithm performance will be called operational data. We
again calculated the identification discrepancy for each fragment, This time it
has been calculated as a difference between the simulated data and the
operational data.  In \cref{fig:eff_result} we present these discrepancies. The
function parameters have been reconstructed with high accuracy, it is not
surprising therefore, that the data classification efficiency is as high as
99.76$\%$. Fragment charge has been properly assigned to all fragments.  550
fragments, out of 228698, were misidentified in mass, all of them being a
species with Z > 4. Compared to the number of fragments misidentified due to noise
superposition, this number rose about 49 and that is a 0.214\textperthousand~of
total number of studied fragments.  
    \begin{figure}[ht]
      \tiny
      \centering
      \begin{minipage}[b]{.45\linewidth}
        \centering
        \includegraphics[keepaspectratio, page=1,width=\textwidth]{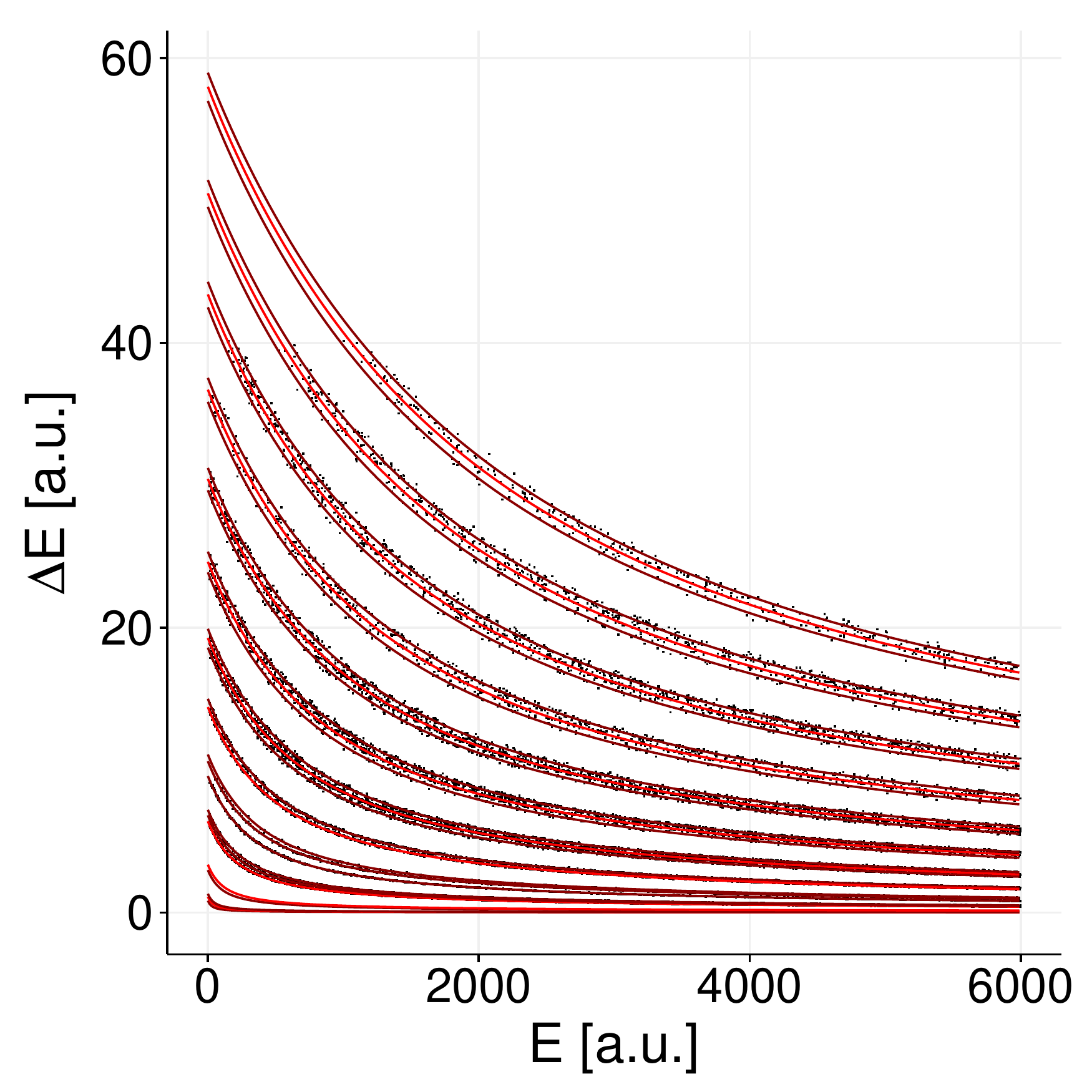}
        \label{fig:withouth}
      \end{minipage}%
      \begin{minipage}[b]{.45\linewidth}
        \centering
        \includegraphics[keepaspectratio, page=1,width=\textwidth]{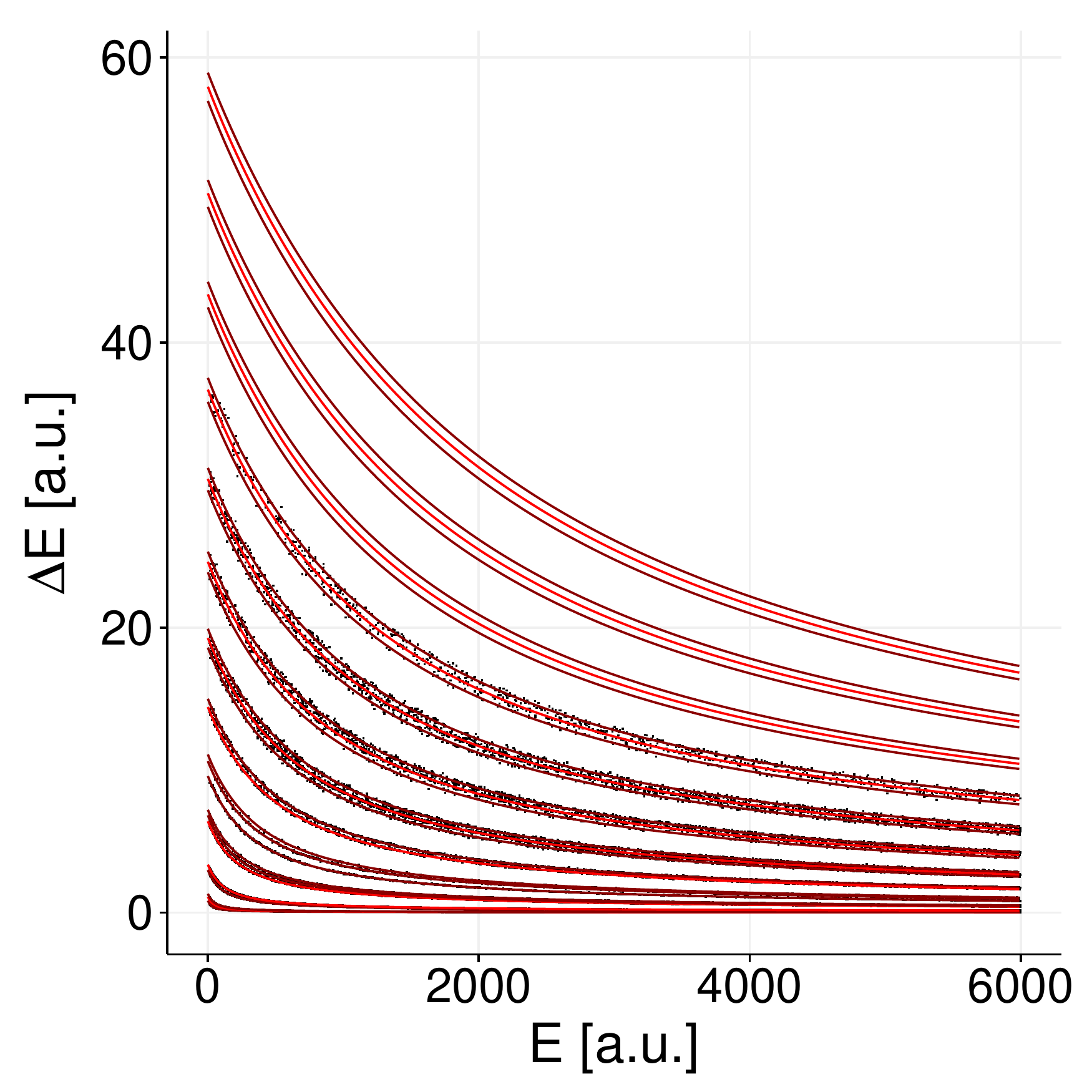}
        \label{fig:withoutmg}
      \end{minipage}
      \caption{Algorithm performance on simulated data with missing isotopes.
      Left: missing H and He isotopes. Right: missing Ne, Na and Mg  isotopes.}
      \label{fig:missingdata}
    \end{figure}
The last test performed was to execute the algorithm on simulated data without some of the isotopes with the
model having number of isotope types not changed ($^{1}_{1}$H up to $^{25}_{12}$Mg). In the left hand side of \cref{fig:missingdata}
the result of algorithm performance without isotopes of H and He is shown, 
in the right hand side of \cref{fig:missingdata} the result for simulated data without Ne, Na and Mg  isotopes. It can be seen
that the algorithm works very well for a such set of data. The conclusion from that test is that in order to 
identify fragments produced in a given experimental reaction it is necessary to create a model encompassing all isotopes expected in
the reaction. Although every telescope detects different number of isotope types,
the algorithm should fit the data properly. 
\section{Summary}
\label{sec:summary}
The goal of this paper was to present the proof of concept of 
using an evolutionary strategy in $\Delta$E - E identification procedure. We have 
simulated labeled data using a model describing $\Delta$E - E correlations. This was followed
by applying our algorithm to that data. In this way we have reconstructed the original 
(simulated) mass (A) and charge (Z) and calculated the efficiency of that reconstruction. 
None of cited articles in section ~\ref{sec:intro} adressed efficiency, since the authors have applied 
their methods directly on experimental, not simulated data.

The method presented in this article shows that the agreement between the
simulated and operational data is 100$\%$ for charge identification.  As
presented in section~\ref{sec:results}, the efficiency of mass identification is
100$\%$ for fragments up to $^{10}$Be. With increasing charge, the distinction
between isotopes decreases, which led to lower mass classification efficiency.
However, only 0.24$\%$ of isotopes were misidentified in mass, which is a
significant result.  It is important to emphasize, that in this paper we have
tested the model describing linear detectors. In the future work we would like
to test the method with various models, including the extended model of
$\Delta$E - E relation, (eq.~\ref{eq:TG2}) for linear detectors and model with
one nonlinear detector (eq. ~\ref{eq:TG2} with eq. ~\ref{eq:Eh} as a
correction). We also would like to test the method sensitivity to various
statistics.  Finally, we plan to apply the method on experimental data collected
by several experiments with various detection arrays.


\appendix

\acknowledgments
We would like to thank dr J. B. Natowitz and dr K. Hagel from Cyclotron Institute,
Texas A$\&$M University for 
improving the quality of this paper by sharing their expertise knowledge.
Furthermore, we would like to thank groups from University of Silesia 
and Jagiellonian University involved in the experiments
on heavy ion reactions at intermediate energies for their kindness and support of this study.


\bibliographystyle{unsrt}
\bibliography{IdentMethod}



%
%
%

\end{document}